\documentclass[aps,prl,amsmath,amssymb,onecolumn,superscriptaddress,showpacs]{revtex4}
\usepackage{amsmath,graphicx,float,epsfig,float,tabularx,multirow}
\usepackage{epstopdf,color,bbm,times}
\usepackage[english]{babel}

\begin{document}
\title{Exploiting the wide dynamic range of Silicon photomultipliers for Quantum Optics applications}
\author{S.~Cassina}
\affiliation{Department of Science and High Technology, University of Insubria, Via Valleggio 11, I-22100 Como (Italy),}
\author{A.~Allevi}
\email{alessia.allevi@uninsubria.it} 
\affiliation{Department of Science and High Technology, University of Insubria, and Institute for Photonics and Nanotechnologies, IFN-CNR, Via Valleggio 11, I-22100 Como (Italy),}
\author{V.~Mascagna}
\affiliation{Department of Science and High Technology, University of Insubria, Via Valleggio 11, I-22100 Como (Italy), and INFN Section of Milano Bicocca, Piazza della Scienza 3, I-20126 Milano (Italy),}
\author{ M.~Prest}
\affiliation{Department of Science and High Technology, University of Insubria, Via Valleggio 11, I-22100 Como (Italy), and INFN Section of Milano Bicocca, Piazza della Scienza 3, I-20126 Milano (Italy),}
\author{E.~Vallazza}
\affiliation{INFN Section of Milano Bicocca, Piazza della Scienza 3, I-20126 Milano (Italy),}
\author{M.~Bondani}
\affiliation{Institute for Photonics and Nanotechnologies, IFN-CNR, Via Valleggio 11, I-22100 Como (Italy).}

\begin{abstract}
Silicon photomultipliers are photon-number-resolving detectors endowed with hundreds of cells enabling them to reveal high-populated quantum optical states. In this paper, we address such a goal by showing the possible acquisition strategies that can be adopted and discussing their advantages and limitations. In particular, we determine the best acquisition solution in order to properly reveal the nature, either classical or nonclassical, of mesoscopic quantum optical states.
\end{abstract}


\maketitle
\section{Introduction} \label{intro}
Silicon photomultipliers (SiPMs) are photon-number resolving detectors characterized by hundreds of pixels (or cells) operated in the Geiger-M$\ddot{\rm u}$ller regime and read in parallel, in order to yield a single output \cite{akindinov,bondarenko,saveliev,piemonte,renker}. By assuming that each cell is fired by at most one photon, the number of fired cells should correspond to the number of impinging photons. However, the non-ideal quantum efficiency and the presence of some drawbacks, such as dark count, optical cross-talk effect and afterpulses, prevent this correspondence. While the efficiency of the detector can be only slightly modified acting on the bias voltage, we have recently demonstrated that it is possible to make the drawbacks negligible by properly acquiring the output of the detector, taking advantage of their different occurence in time \cite{scirep19}. Moreover, in Ref.~\cite{OL19} we have investigated in which way the drawbacks affect the observation of nonclassical correlations between the two parties of a multi-mode twin-beam state.\\
In this paper, we focus on the temporal development of the output signal in order to reduce the spurious contributions to the final signal, and to select only the information on the light. To do this, we follow two approaches: 1) we use the minimum integration gate; 2) we select the peak values and check the quality of the extracted information by calculating a relevant parameter (the noise reduction factor) for both classically- and nonclassically-correlated light states. To this aim, we consider and compare different kinds of amplifiers and digitizers used to sample the detector output and show the advantages and limitations of all the employed devices. Through this analysis, we also study the limits imposed by nonlinearities and saturation effects of some parts of the acquisition chain, having in mind the exploitation of the high dynamic range of SiPMs to detect well-populated states of light. Indeed, reliably detecting the number of photons in every pulse of highly-populated states is the key resource to implement homodyne-like schemes with a mesoscopic local oscillator (up to 50 mean photon numbers), as required to achieve an optimal quantum state reconstruction \cite{NJP19,PLA20}. Increasing the dynamic range is also required if the considered states of light are characterized by large fluctuations, such as in the case of superthermal states of light \cite{OL15,qmetro17}. Moreover, since the ultimate aim of our research activity is the exploitation of SiPMs for the reconstruction of nonclassical states of light, in our work we do not limit ourselves to the reconstruction of statistical properties, but we also analyze the shot-by-shot measurements, which are the main ingredient of the calculation of photon-number correlations and nonclassicality criteria.
\section{Methods} \label{sec:methods}
\begin{figure}
\resizebox{0.45\textwidth}{!}{\includegraphics{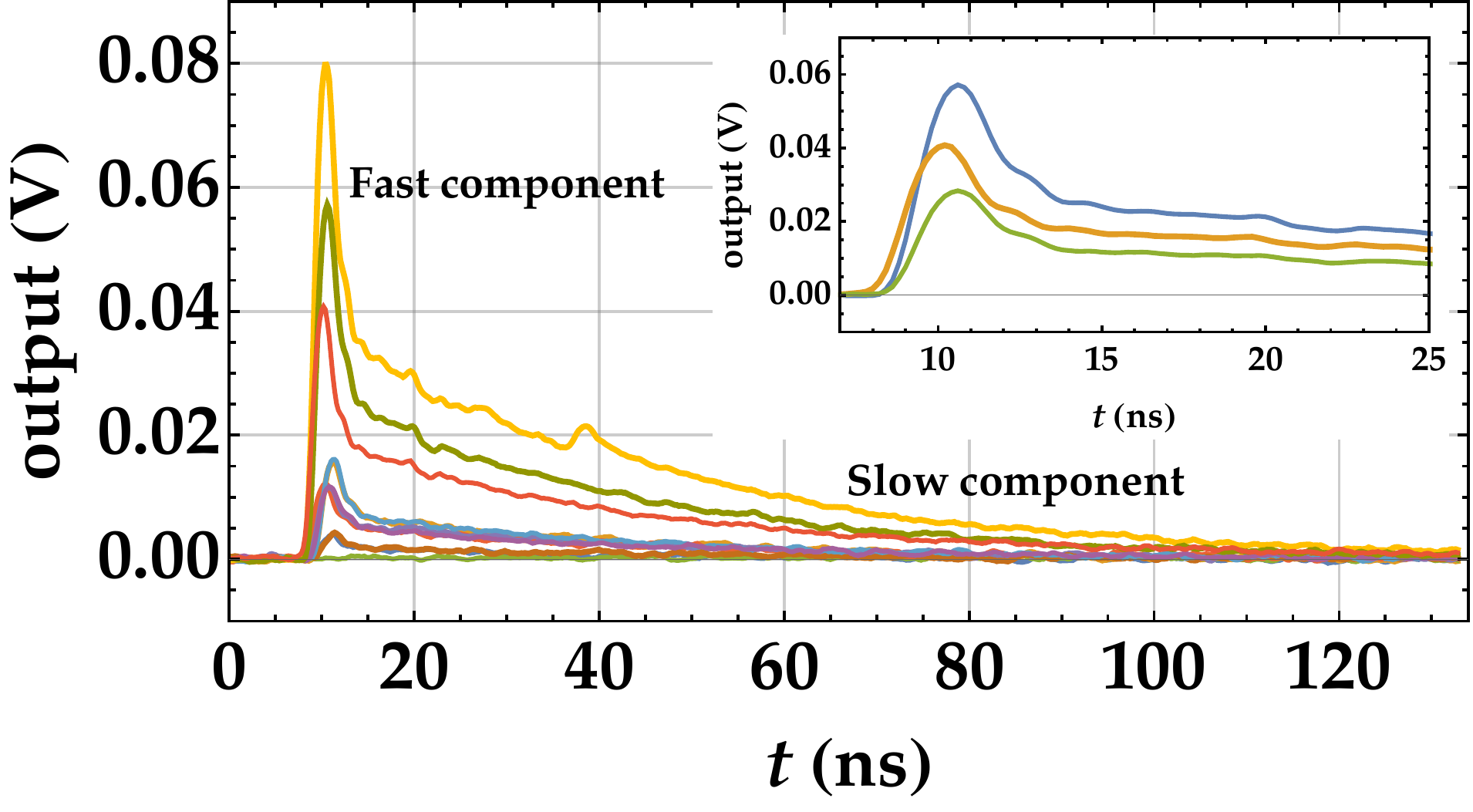}}
\caption{Main: Typical single-shot detector outputs of MPPC S13360-1350CS SiPMs. 
Inset: Zoom of the fast component of some of the waveforms.}
\label{fig:signal}
\end{figure}

In Fig.~\ref{fig:signal} we plot a number of single-shot detector outputs of the S13360 SiPM series of Hamamatsu. This SiPM series is characterized by  low values of cross talk and dark counts and negligible afterpulse probability. 
The rising edge of the signal, corresponding to the charge process, is very fast, lasting less than 1~ns, while the falling edge, corresponding to the discharge process, is much longer (hundreds of ns depending on the specific model).
We assume that the output charge, given by the sum of the signals from all the fired cells, is proportional to the number of detected photons, modified by the presence of dark counts and cross talk. The value of the output charge is given by the integral of the output signal, that is the area under the curves in Fig.~\ref{fig:signal}. Since the presence of cross-talk effects manifests at delayed times with respect to the main detection peak, limiting the integrated area reduces such spurious contributions.
The main question starting our analysis is thus if even a portion of the area or the peak height alone contain the same information on light as the entire area.
Indeed, in some previous works of ours \cite{scirep19,OL19}, we have demonstrated that integrating the signal over a gate shorter than the entire curve reduces the incidence of spurious effects and makes it possible the observation of the nonclassical character of quantum states of light. However, to understand if such a procedure has general value, one should model the signal output.
In Refs.~\cite{corsi,seifert} the rising edge of the output is described as a single exponential or, more properly, by two exponentials with the same time constant, while the falling edge is modelled as the sum of two exponentials with distinct decay times. This means that, formally, there is a proportionality between the area under the rising edge and the height of the peak. On the contrary, no perfect proportionality between the total area and a portion of the area under the falling edge is expected. 
Thus, to successfully exploit SiPMs for applications, such as for the realization of a homodyne-like detection scheme \cite{PLA20}, it is important that the signal output is properly acquired and analyzed.
All these observations led us to test different detection chains based on different devices and to compare their performance. In the following, we describe all the investigated acquisition chains by emphasizing their advantages and limitations with respect to the above-mentioned goal, that is the detection of well-populated states of light.\\
\subsection{The sensors} \label{sec:sensors}
The SiPMs we used are the MPPC S13360-1350CS produced by Hamamatsu Photonics \cite{hama}. They consist of 667 pixels in a 1.3 $\times$ 1.3 mm$^2$ photosensitive area, with a pixel pitch equal to 50$~\mu$m and a maximum quantum efficiency of 40$\%$ at 460 nm \cite{piemonte,frach}. In our experiment we operated at 523 nm, where the quantum efficiency is still good ($\sim 38\%$).
As anticipated, the main drawbacks that affect such detectors are given by the dark counts, the optical cross talk and the afterpulses.
Dark counts consist of spurious avalanches triggered by thermally-generated charge carriers \cite{dinu,akiba}, whereas cross-talk effect is due to the spontaneous emission of secondary infrared photons inside the silicon substrate after an avalanche process is generated in a cell by a detection event. The secondary photons can be detected by another neighboring cell, generating the same kind of signal as the primary photons \cite{lacaita,buzhan,du,gola}.
This effect can be simultaneous to the light signal (prompt cross talk) or retarded (delayed cross talk) \cite{nagy}. While the contribution of the delayed cross talk to the output signal can be removed by reducing the integration gate, the prompt one is indistinguishable from the light signal. Finally, afterpulses are avalanches produced by the photoelectrons captured by the geometric imperfections of the device structure and released at a later time with respect to the light signal \cite{cova}.
The model of SiPMs we considered is endowed with a moderate dark-count rate (the typical value reported in the datasheet is $\sim$ 90 kHz), a low cross-talk probability ($\sim$3 $\%$) and a negligible afterpulse probability (less than 1 $\%$).
\subsection{The amplifiers} \label{sec:amplifiers}
\begin{figure}
\centering
\resizebox{0.45\textwidth}{!}{\includegraphics{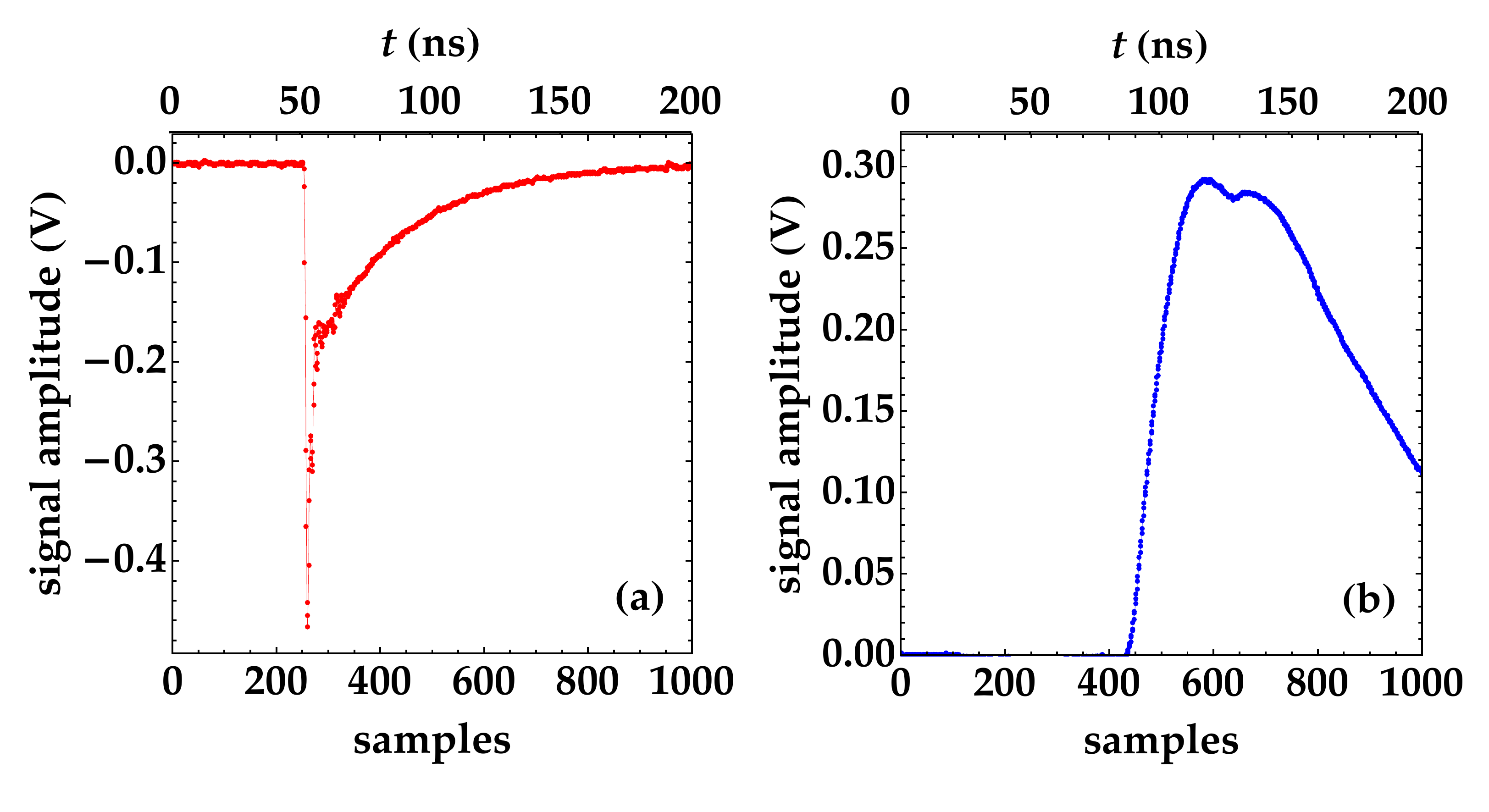}}
\caption{(a): typical output of the Caen PSAU amplifier. (b): typical output of the slow amplifier. Both signal outputs have been sampled at 5~GS/s by a DRS4 digitizer (see text).}
\label{fig:ampli}
\end{figure}

In general, the SiPM output is externally amplified. We consider amplifiers of two different kinds.
The first device is a fast inverting amplifier embedded in the computer-based Caen SP5600 Power Supply and Amplification Unit (PSAU) \cite{caen_ampli}. Once amplified, the output temporal shape is essentially the same as that of the SiPM. Such an amplifier is quite versatile since its gain can be changed from 1 dB up to 40 dB in unit steps.\\
The second amplifier is a home-made circuit including a slow non-inverting amplifier with two amplifying stages \cite{opamp}, each one having a gain of 5.5 for a total gain of 29.6~dB \cite{AD828}. As detailed in the following, this second choice allowed us to better select the peak of the detector output since it stretches the rising edge from less than 1~ns to 50-ns.
For a fair comparison, in Fig.~\ref{fig:ampli} we show two typical outputs of the two amplifiers, both digitized with the DRS4 digitizer described below. Panel (a) displays the output of the Caen amplifier, while panel (b) that of the slow amplifier.
By observing the two different temporal behaviors, we can assess that the fast amplifier is more indicated to collect the entire output, while the slow one is more suitable for catching the peak, as it will appear clearer in the next Section.\\
\subsection{The digitizers} \label{sec:digitalization}
\begin{figure}
\centering
\resizebox{0.45\textwidth}{!}{\includegraphics{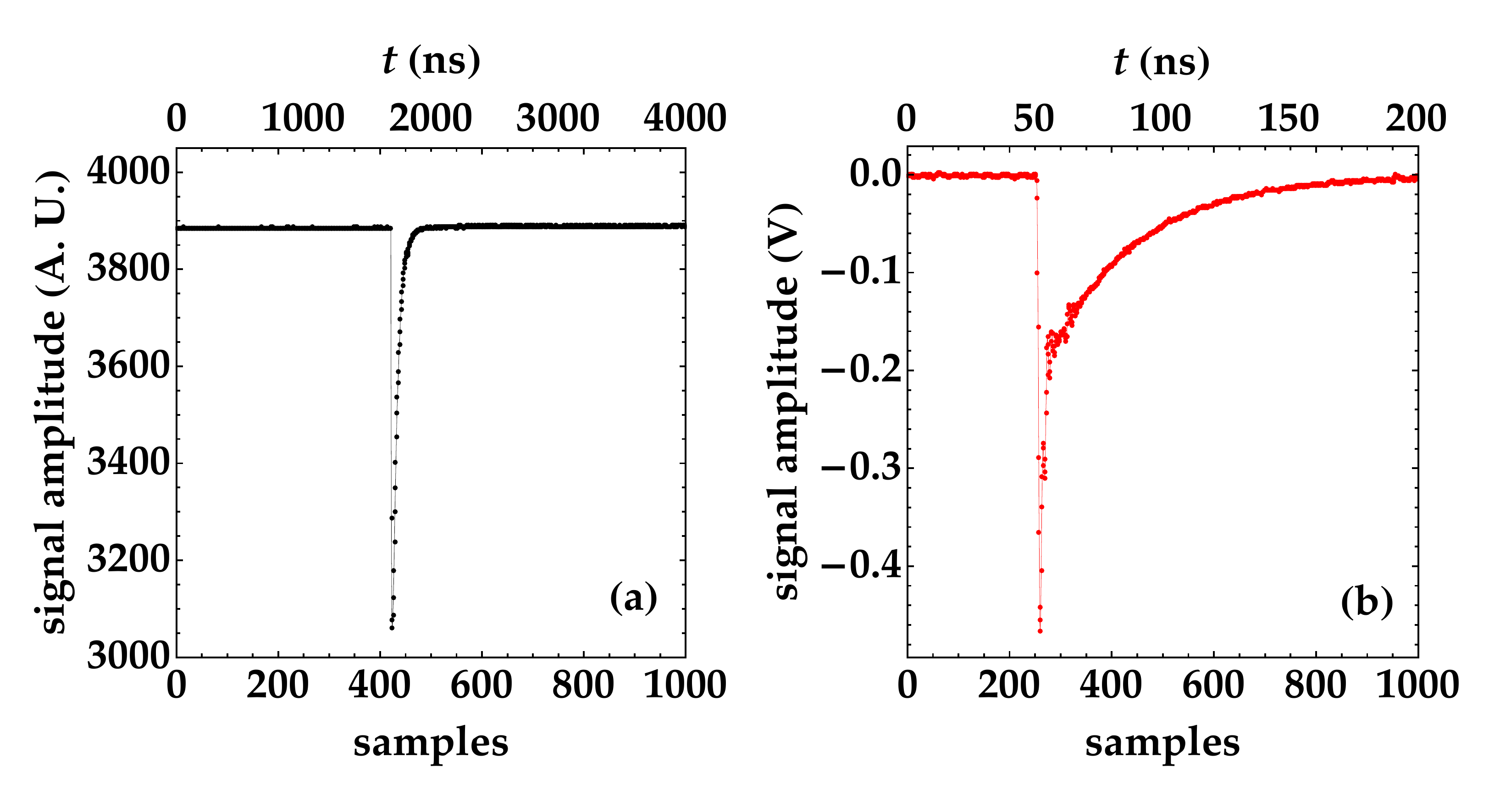}}
\caption{(a): typical output of a Caen-amplified signal digitized at 250MS/s by the Caen digitizer (PSAU+DT5720). (b): typical output of a Caen-amplified signal digitized at 5~GS/s by the DRS4 digitizer (PSAU+DRS4). The signal traces are negative because of the Caen amplifier.}
\label{fig:digi}
\end{figure}

To acquire information from the signal trace, two possible strategies can be implemented: an analogical integration or a signal digitalization followed by an offline integration. In Ref.~\cite{OL19}, the best results in terms of nonclassicality detection were achieved by amplifying the signals with a PSAU unit and integrating them with boxcar-gated integrators over a very short gate (10-ns long). However, as we remarked in that work, this procedure can be fragile when the integration gate is short, as it gives no direct control on the unwanted presence of electronic signal jitter.\\
To avoid the problems of analogical integration, in this work we decided to proceed with an offline analysis of the digitized amplified traces. We used two different models of digitizers: the first one is the computer-based Caen DT5720 desktop waveform digitizer, a two-channel device endowed with 12-bit resolution, a full scale range of 2 V peak-to-peak, and a sampling rate ranging from 31.25 to 250 MS/s \cite{caen_digi}. Since its minimum sampling time is 4~ns, this device is suitable for the acquisition of rather long signals, such as the entire amplified output. The second digitizer we used is the model DRS4 produced by the Paul Scherrer Institute \cite{Ritt,drs4}. Also this device has 12-bit resolution, whereas its peak-to-peak voltage range is limited to 1 V, but its sampling frequency can be changed from 1 up to 5 GS/s. Two typical signals, both amplified by the PSAU unit and digitized by the devices described above, are shown in the two panels of Fig.~\ref{fig:digi}.
By inspecting the two panels, it clearly appears that the signal acquired with the Caen digitizer lacks in most details characterizing the signal acquired with DRS4 digitizer. Moreover, the resulting shape of the peak is completely different, indicating an undersampling of the trace that does not allow following the fast part of the SiPM signal.
\subsection{The light sources} \label{sec:lightsources}
In order to test the different acquisition chains described in the previous Sections, we generated different classical and quantum optical states.
The light source is a mode-locked Nd:YLF laser regeneratively amplified at 500 Hz, emitting the fundamental beam at 1047 nm, the second harmonic at 523 nm and the third-harmonic at 349 nm. In the following Section, we show the results obtained by considering three kinds of optical states: coherent states, pseudo-thermal states and multi-mode twin-beam states.
\begin{figure}
\resizebox{0.49\textwidth}{!}{\includegraphics{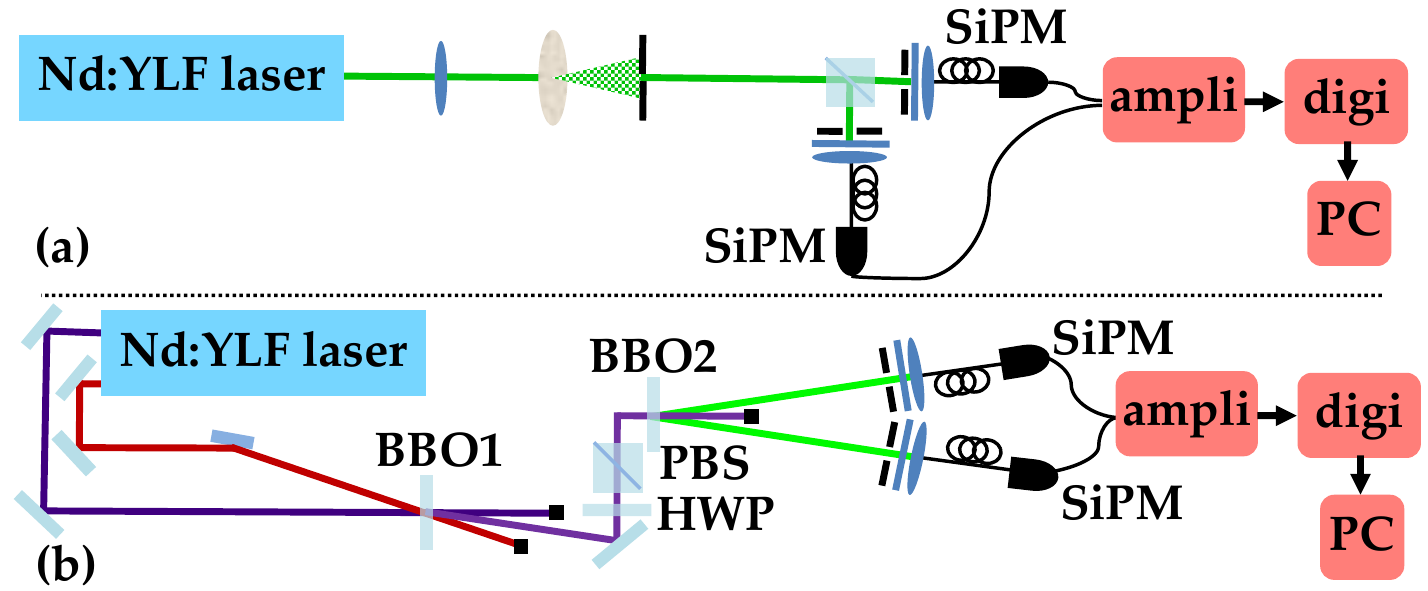}}
\caption{Sketch of the experimental setup. (a): Scheme used to produce and detect single-mode pseudo-thermal states. (b): Scheme used to produce and detect multi-mode twin-beam states. See the text for details.}
\label{fig:setup}
\end{figure}
To produce the classical states (coherent and pseudo thermal), we exploited the second-harmonic of the laser to match the sensitivity region of SiPMs. The coherent state was obtained by taking a portion of the laser light, while a rotating ground glass disk combined with a pin-hole to select a single speckle was used to generate pseudo-thermal light (see Fig.~\ref{fig:setup}(a)). The light was then split into two beams by means of an adjustable beam splitter (BS) made of a polarizing beam splitter preceded by a half-wave plate allowing a careful balancing of the mean values in the two beams. At the two outputs of the BS, two achromatic doublets focused the two beams into multi-mode fibers (1-mm core diameter) to deliver them to SiPMs. A set of neutral density filters was used in front of the BS to change the mean value of the light.\\
The quantum states were twin-beam states generated by pumping a $\beta$-barium-borate crystal (BBO2 in Fig.~\ref{fig:setup}(b)) with the fourth harmonic of the laser (at 262 nm, 3.5-ps pulse duration).
We selected two portions of the generated twin beam at frequency degeneracy (523 nm) both in space (by means of irises 7-mm wide) and in spectrum (by means of bandpass filters 10-nm wide) and sent each of them to a SiPM by focusing it into a multi-mode fiber (1-mm core diameter) with an achromatic doublet. A half-wave plate and a polarizing beam splitter were used to change the pump energy and, consequently, the mean number of photons of the twin beam.
\section{Results} \label{sec:results}
\subsection{Pulse-height spectra}\label{sec:phs}
\begin{figure}
\resizebox{0.45\textwidth}{!}{\includegraphics{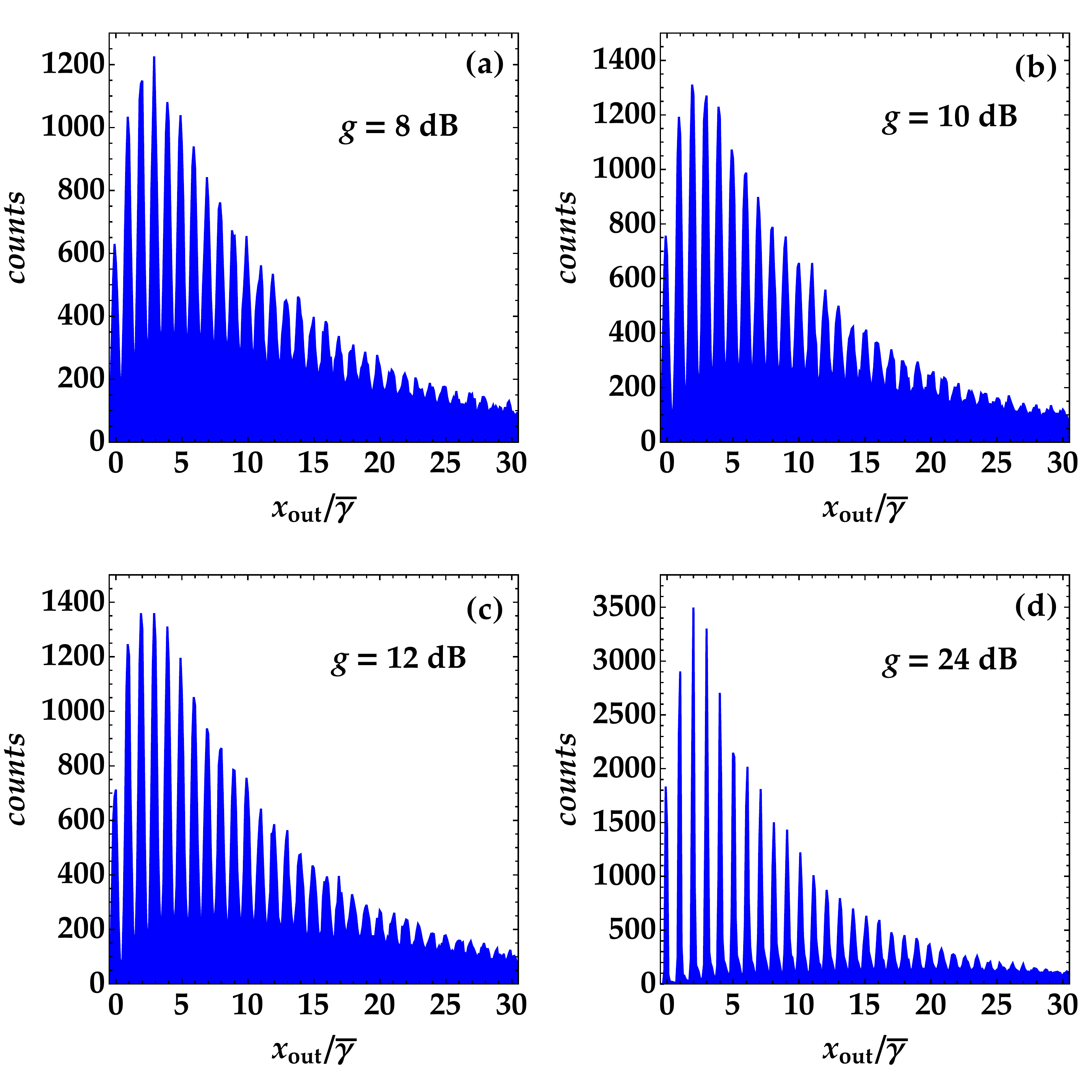}}
\caption{Pulse-height spectra for a pseudo-thermal state obtained with the system PSAU+DRS4 and integrated over $\tau=70$~ns for the different amplifier gains $g$ indicated in the figure panels. The mean value of the pseudo-thermal state is $x_{\rm out}/\bar{\gamma}\sim 14$. The corresponding visibility values are (a): $v = 0.39 \pm 0.04$, (b): $v = 0.43 \pm 0.05$, (c): $v = 0.49 \pm 0.04$ and (d): $v = 0.70 \pm 0.06$.}
\label{fig:phsgain}
\end{figure}

\begin{figure}
\resizebox{0.45\textwidth}{!}{\includegraphics{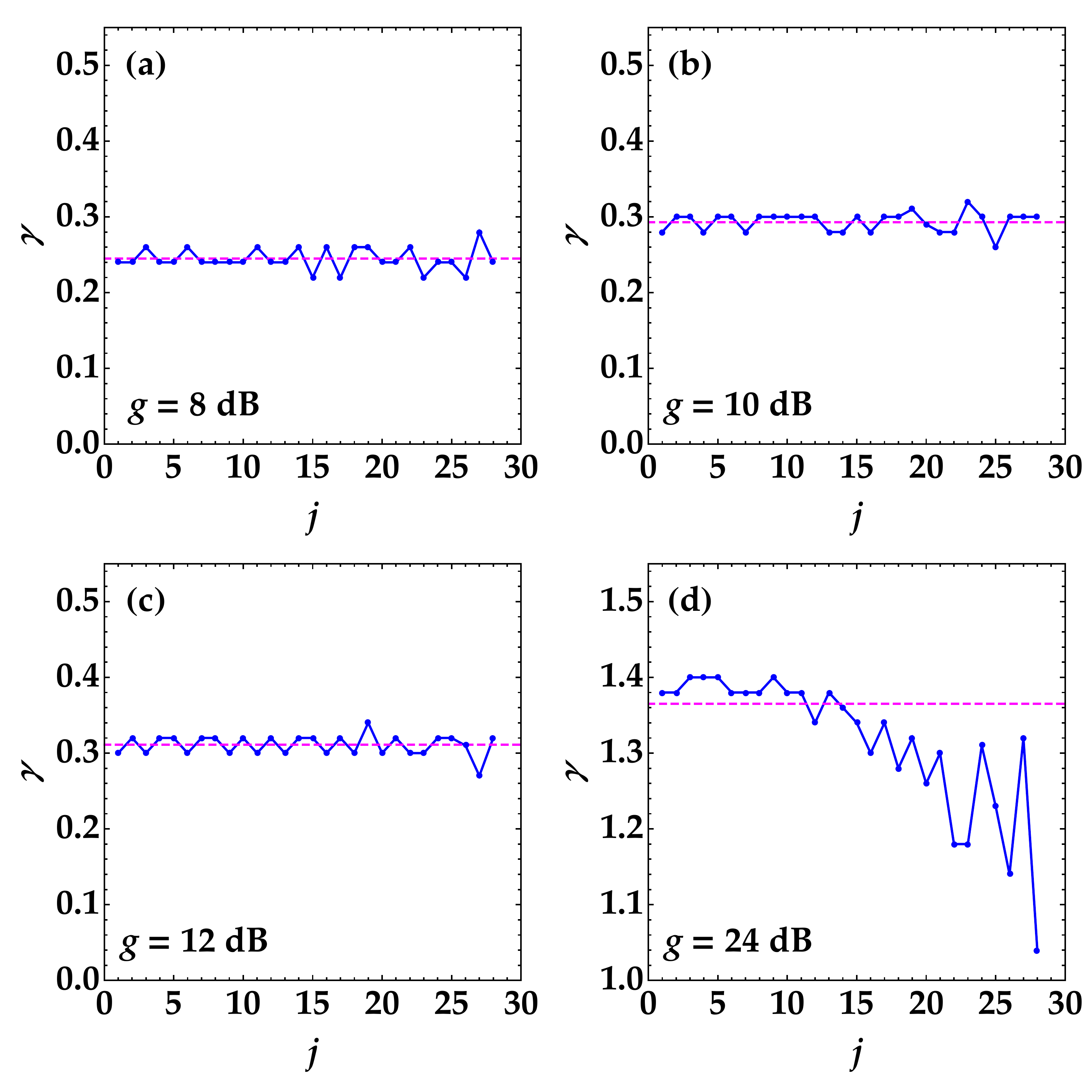}}
\caption{Plot of the peak-to-peak distance in the pulse-height spectra as a function of the peak number for the data in Fig.~\ref{fig:phsgain} (blue lines). The dashed magenta line represents $\bar{\gamma}$, the mean peak-to-peak distance used to calibrate the abscissa in all pulse-height spectra figures.}
\label{fig:gamma}
\end{figure}

\begin{figure}[t]
\resizebox{0.45\textwidth}{!}{\includegraphics{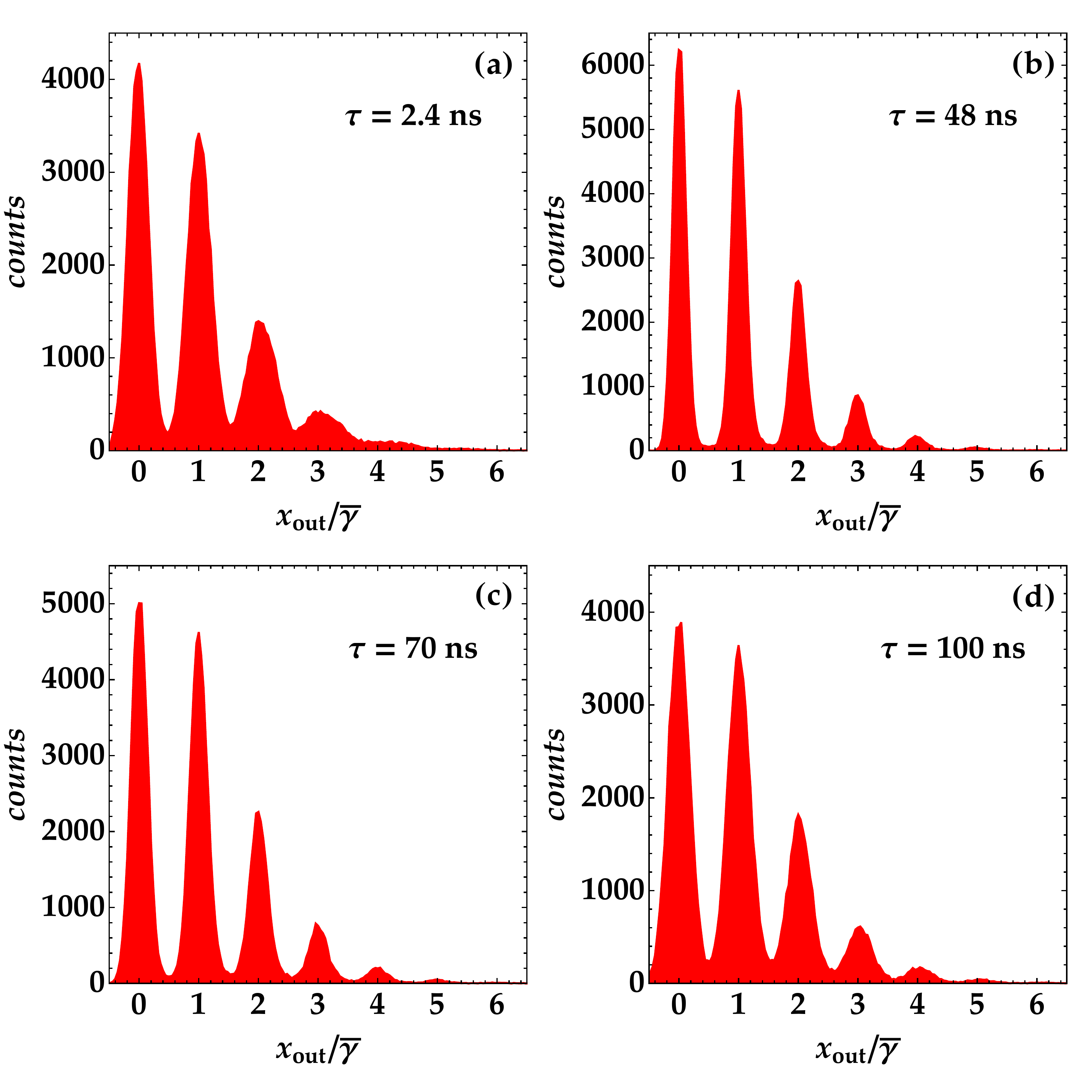}}
\caption{Pulse-height spectra for a coherent state obtained with the system PSAU+DRS4 at the gain g~$= 12$~dB for the different integration gate widths indicated in the figure panels. The mean value of the detected coherent state is $\langle x_{\rm out}\rangle/\bar{\gamma}\sim 1$. The corresponding visibility values are $v = 0.74 \pm 0.07$ for $\tau = 2.4$~ns, $v = 0.97 \pm 0.01$ for $\tau = 48$~ns, $v = 0.95 \pm 0.01$ for $\tau = 70$~ns, and $v = 0.88 \pm 0.01$ for $\tau = 100$~ns.}
\label{fig:phsmis14}
\end{figure}

\begin{figure}[t]
\resizebox{0.45\textwidth}{!}{\includegraphics{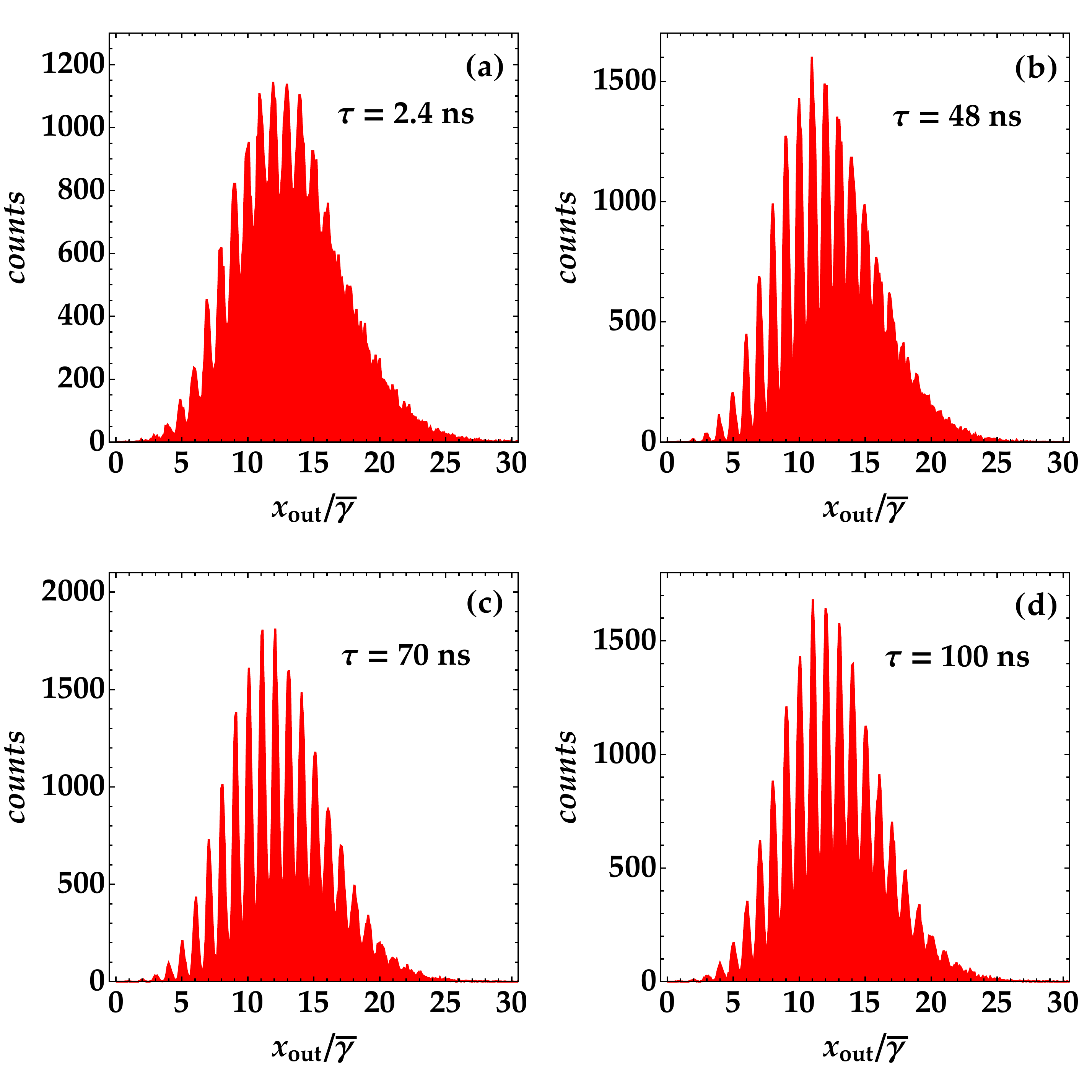}}
\caption{Same as in Fig.~\ref{fig:phsmis14} for a coherent state having mean value $\langle x_{\rm out}\rangle/\bar{\gamma}\sim 13$. The corresponding visibility values are $v = 0.32 \pm 0.05$ for $\tau = 2.4$~ns, $v = 0.60 \pm 0.06$ for $\tau = 48$~ns, $v = 0.71 \pm 0.04$ for $\tau = 70$~ns, and $v = 0.63 \pm 0.04$ for $\tau = 100$~ns.}
\label{fig:phsmis21}
\end{figure}

First of all, we investigate the reconstruction of the statistical properties of some optical states. 
SiPMs are photon-number-resolving detectors since their pulse-height spectrum is characterized by a multi-peak structure \cite{peaks}. Upon a proper normalization procedure, each peak of the spectrum of PNR detectors can be interpreted as the probability that a given number of photons is detected \cite{JMO,PNR14}. Here, we want to compare the pulse-height spectra obtained by measuring a given state and integrating the detector output over either different amplifier gains or different gate widths.
In Fig.~\ref{fig:phsgain}, we compare the pulse-height spectra of a pseudo-thermal state plotted as a function of $x_{\rm out}/ \bar{\gamma}$, $x_{\rm out}$ being the detection output and $\bar{\gamma}$ the mean peak-to-peak distance, representing the gain of the detection chain. As extensively discussed in previous papers of ours (see $e.g.$ Ref.~\cite{PNR14}), $\bar{\gamma}$ includes all the amplification stages of the detection apparatus, and is assumed to be sharp enough \cite{PRA09}. Note that $\bar{\gamma}$ can be determined in different ways: by the self-consistent method explained in Ref.~\cite{JMO}, by a multi-Gaussian fitting procedure, as described in Ref.~\cite{JOSABramilli}, or as the mean distance between consecutive peaks of the pulse-height spectrum \cite{PNR14}.\\
The mean value of the state in Fig.~\ref{fig:phsgain} is $x_{\rm out}/\bar{\gamma}\sim 14$, measured combining the amplifier of the PSAU Unit and the digitizer DRS4 operated at 5~GS/s. The gain of the amplifier was varied, while the offline integration gate was kept fixed at $\tau=70$~ns.
We note that, as expected, the quality of the spectra seems to improve at increasing gain values, and the noise between neighboring peaks visibly decreases. In order to quantify the quality of the spectra, we define the visibility as
\begin{equation}
v = \sum_{i=1}^N \frac{(M_i - m_i)}{N(M_i + m_i)} ,
\end{equation}
in which $i$ is the i-th peak of the pulse-height spectrum, $N$ is the total number of visible peaks, whereas $M_i$ and $m_i$ are the values of the i-th peak height and its consecutive valley, respectively. For the plots in the figure we obtained $v = 0.39 \pm 0.04$ for 8-dB gain, $v = 0.43 \pm 0.05$ for 10-dB gain, $v = 0.49 \pm 0.04$ for 12-dB gain, and $v = 0.70 \pm 0.06$ for 24-dB gain. All these values prove the qualitative impression about the resolution. Indeed, for the low gain values the results are rather similar, while for the highest gain value $v$ is definitely larger, even if in this last case the peaks corresponding to large values of $x_{\rm out}/ \bar{\gamma}$ appear less resolved probably because of some saturation effects of the amplifier. To check the presence of saturation, in Fig.~\ref{fig:gamma} we plot the peak-to-peak distance, $\gamma$, which represents the overall gain of the detection chain, as a function of the peak number. For a perfectly linear system, $\gamma$ should be constant. In Fig.~\ref{fig:gamma} we observe that $\gamma$ is constant for the lower gain values (panels (a)-(c)), while for the highest amplifier gain $\gamma$ dramatically decreases after about 10 peaks (panel (d)) due to amplifier saturation. This implies that detected-photon values exceeding 10 cannot be reliably recovered. \\
\noindent
To avoid saturation issues, in the following we consider pulse-height spectra obtained with gain values limited to 12~dB, which represent a good compromise between peak resolution and amplifier linearity.\\
In Fig.s~\ref{fig:phsmis14} and \ref{fig:phsmis21} we show the results achieved when the SiPM outputs are amplified by the PSAU Unit ($g=12$~dB) and then digitized by the DRS4 digitizer operated at 5~GS/s. The mean value of the reconstructed state is $x_{\rm out}/\bar{\gamma}\sim 1$ in Fig.~\ref{fig:phsmis14} and $x_{\rm out}/\bar{\gamma}\sim 13$ in Fig.~\ref{fig:phsmis21}. In both cases, the offline integration gate widths are $\tau=2.4$, 48, 70, and 100~ns.
By comparing the two multipanel figures, it is clear that the best reconstructed pulse-height spectrum is differently achieved in the two cases. Indeed, for $x_{\rm out}/\bar{\gamma}\sim 1$, the best performing gate width seems to be $\tau = 48$~ns. This impression is quantitatively proved by the values of $v$, which are $0.74 \pm 0.07$, $0.97 \pm 0.01$, $0.95 \pm 0.01$, $0.88 \pm 0.01$, respectively.
On the contrary, for what concerns Fig.~\ref{fig:phsmis21}, the shorter the gate the less resolved the tail of the spectrum: the obtained values of $v$ are equal to $0.32 \pm 0.05$, $0.60 \pm 0.06$, $0.71 \pm 0.04$, $0.63 \pm 0.04$, respectively.\\
The comparison performed so far demonstrates that reducing the integration gate is not the correct strategy at increasing mean number of photons. This observation is a clear evidence of what we have already noticed in Section~\ref{sec:methods}: when the light state is quite populated, the integral of just a part of the falling edge of the output signal is not proportional to the integral of the whole output.
\\
In general, the plots in Fig.s~\ref{fig:phsmis14} and \ref{fig:phsmis21} prove that, when the mean value of the light is quite large, integrating the signal output of the detection chain can be critical. In particular, it seems that the best resolution is achieved by considering the entire output trace, even if this choice entails the presence of delayed spurious effects.

\begin{figure}[h!]
\resizebox{0.45\textwidth}{!}{\includegraphics{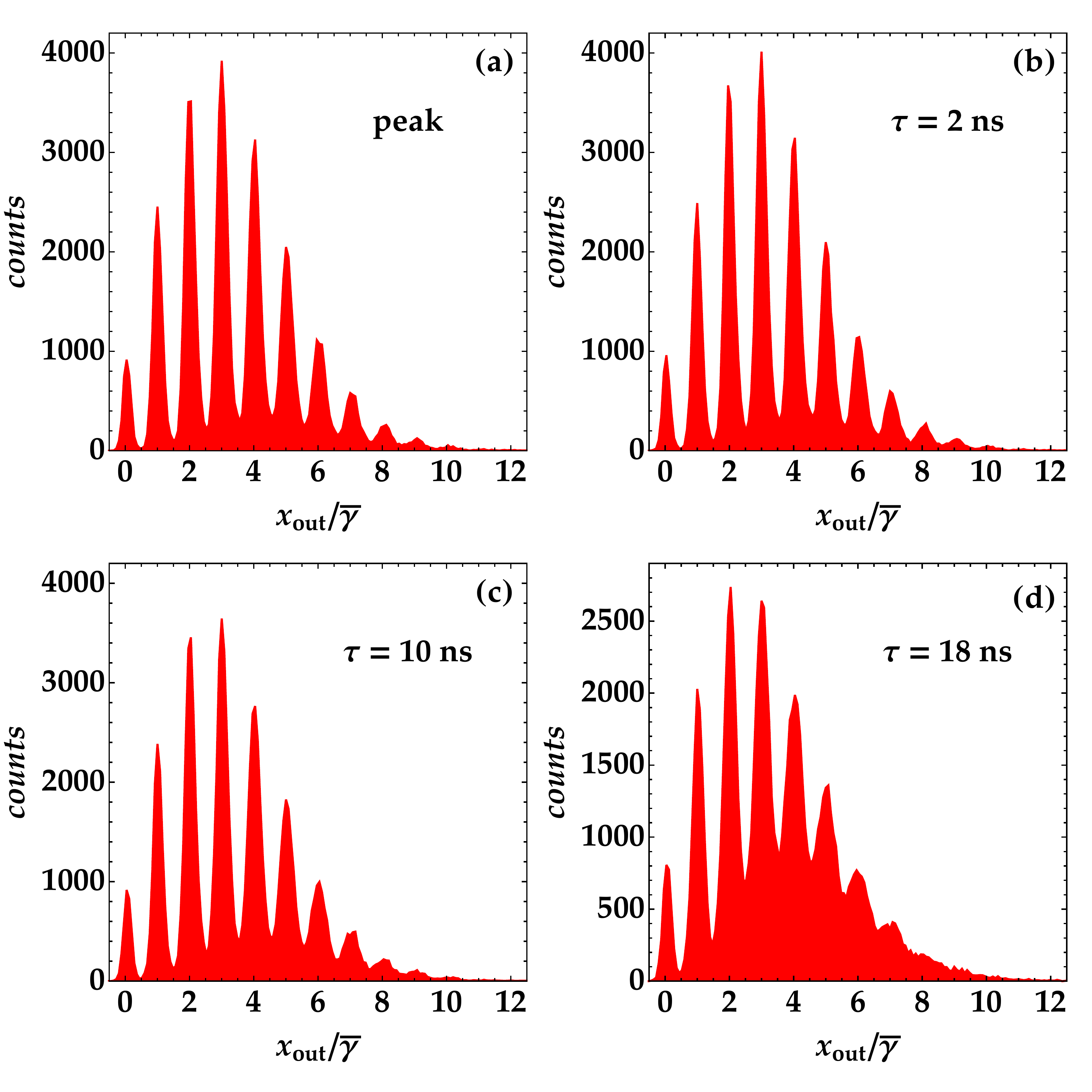}}
\caption{Pulse-height spectra for a coherent state obtained with the system slow amplifier+DRS4 obtained either by selecting the peak value or integrating the rising edge of the signal over different integration widths as displayed in the figure panels. The mean value of the detected coherent state is $\langle x_{\rm out}\rangle/\bar{\gamma}\sim 3.6$. The corresponding visibility values are $v = 0.82 \pm 0.06$ for the peak value, $v = 0.82 \pm 0.06$ for $\tau = 2$~ns, $v = 0.75 \pm 0.08$ for $\tau = 10$~ns, and $v = 0.57 \pm 0.11$ for $\tau = 18$~ns.}
\label{fig:phsDRS4}
\end{figure}

\noindent
We thus consider a different strategy and substitute the fast PSAU amplifier with a slow amplifier and digitize the amplified signal with the DRS4 at 5~GS/s. We then integrate a portion of the rising edge or select the peak value. The data are presented in Fig.~\ref{fig:phsDRS4}.
The best resolved spectrum is that obtained from the peak values for which $v = 0.82 \pm 0.06$. Such a value coincides with that from the integration over $\tau=2$~ns. On the contrary, for larger integration widths the spectra are noisier ($v = 0.75 \pm 0.08$ for $\tau=10$~ns and $v = 0.57 \pm 0.11$ for $\tau=18$~ns), even if definitely better than those in Fig.s~\ref{fig:phsgain} - \ref{fig:phsmis14}. \\
The analysis performed so far proves that the proper reconstruction of the light statistics depends on the integration gate width and that either the entire trace signal or the peak values should be considered. Obviously, the latter choice requires a proper shaping of the detector output, but offers the advantage of avoiding the main drawbacks of SiPMs.

\subsection{Classical correlations}\label{sec:pseudo-thermal}
For Quantum Optics applications, the reconstruction of the statistical properties is not enough to guarantee that the information about the light state under examination is properly acquired. In particular, in many situations, such as in state-discrimination protocols \cite{OLcorr,OE17}, it is crucial that each single pulse is correctly detected. Hereafter we compare the different acquisition strategies discussed above for the calculation of shot-by-shot photon-number correlations and nonclassicality criteria.
As the experimental estimator we consider the noise reduction factor, a quantity usually employed to quantify the nonclassical character of quantum-correlated bipartite states of light, such as twin-beam states. The noise reduction factor is defined as
\begin{equation}\label{eq:Rphot}
R= \frac{\sigma^2(n_1-n_2)}{\langle n_1\rangle + \langle n_2 \rangle}
\end{equation}
$n_j$ being the number of photons in the two components of the bipartite state.
According to the definition, $R$ is the ratio between the variance of the photon-number difference at the two outputs and the shot-noise level.
It is well-known that $R$ must be less than 1 in case of quantum-correlated optical states, while it is identically equal to 1 in case of classical states of light \cite{jointdiff}. Thus, we expect that for pseudo-thermal states, both single-mode and multi-mode, the noise reduction factor is unitary. Actually, the value of $R$ that can be experimentally measured is affected by non-unit quantum efficiency, dark counts, cross-talk and imbalance between the components of the bipartite state. For this reason, the experimental value of $R$ is
\begin{equation}\label{eq:Rexp}
R= \frac{\sigma^2(k_1-k_2)}{\langle k_1\rangle + \langle k_2 \rangle}
\end{equation}
$k_j$ being the number of detected photons including all the experimental effects.

\begin{figure}
\resizebox{0.45\textwidth}{!}{\includegraphics{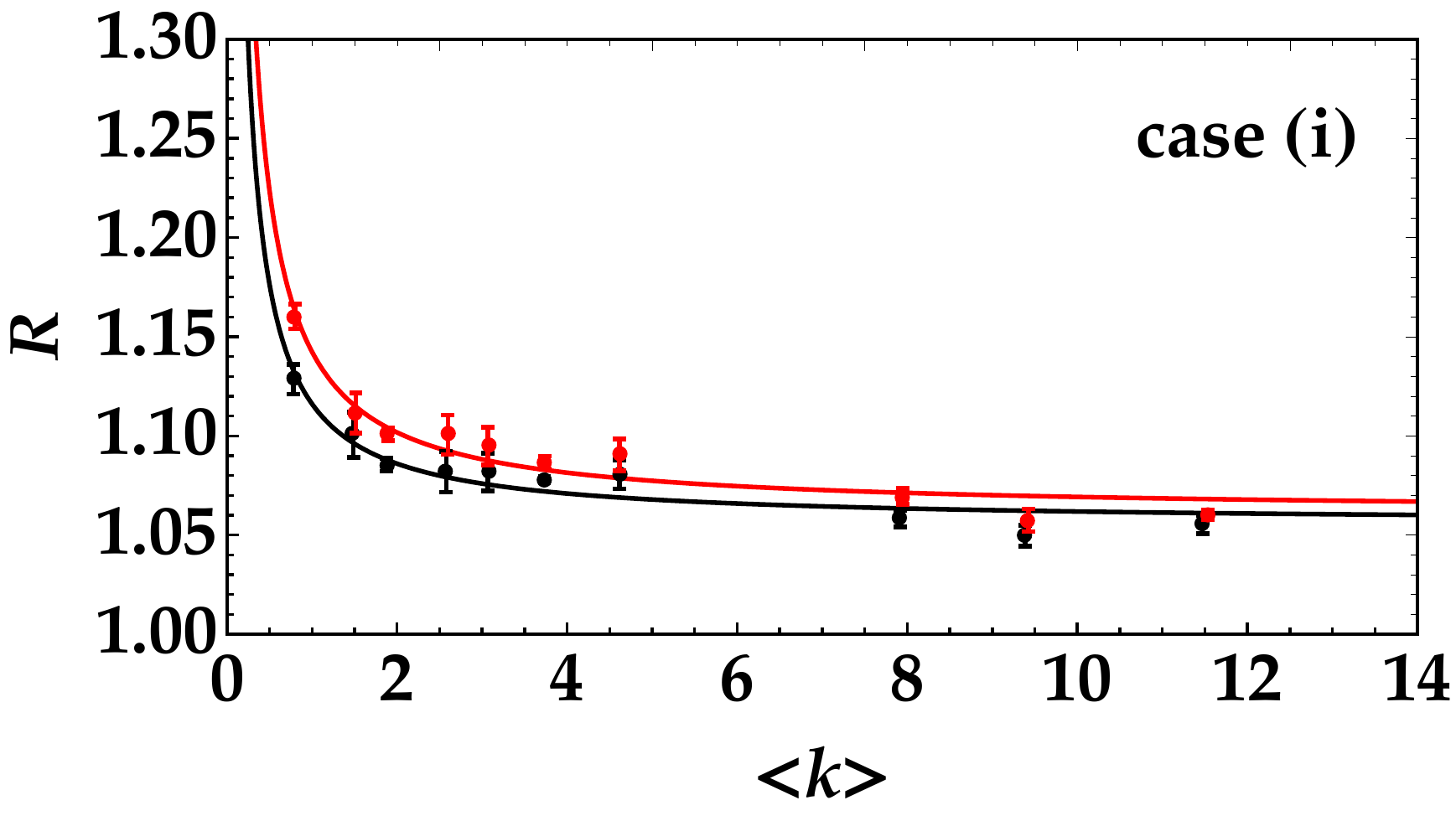}}
\caption{Noise reduction factor for the single-mode thermal state as a function of the mean number of photons detected at a BS output. Colored dots and error bars: experimental data obtained with the CAEN system PSAU ($g = 12$~dB) and DT5720 digitizer integrating the traces over $\tau=48$~ns (black dots) and $\tau =80$~ns (red dots); colored lines: theoretical expectations according to Eq.~(\ref{eq:RtotTWBt}). The fitting parameters for $\tau=48$~ns are: $\epsilon = 0.029$, $\langle m_{\rm 1dc}\rangle = 0.231$, $\langle m_{\rm 2dc}\rangle = 0.569$, $t=0.999$ and $\chi^{(2)}=4.8$, while those for $\tau = 80$~ns are: $\epsilon = 0.031$, $\langle m_{\rm 1dc}\rangle= 0.205$, $\langle m_{\rm 2dc}\rangle = 0.595$, $t=0.999$ and $\chi^{(2)}=3.6$.}
\label{R_therm10jan}
\end{figure}

\begin{figure}
\resizebox{0.45\textwidth}{!}{\includegraphics{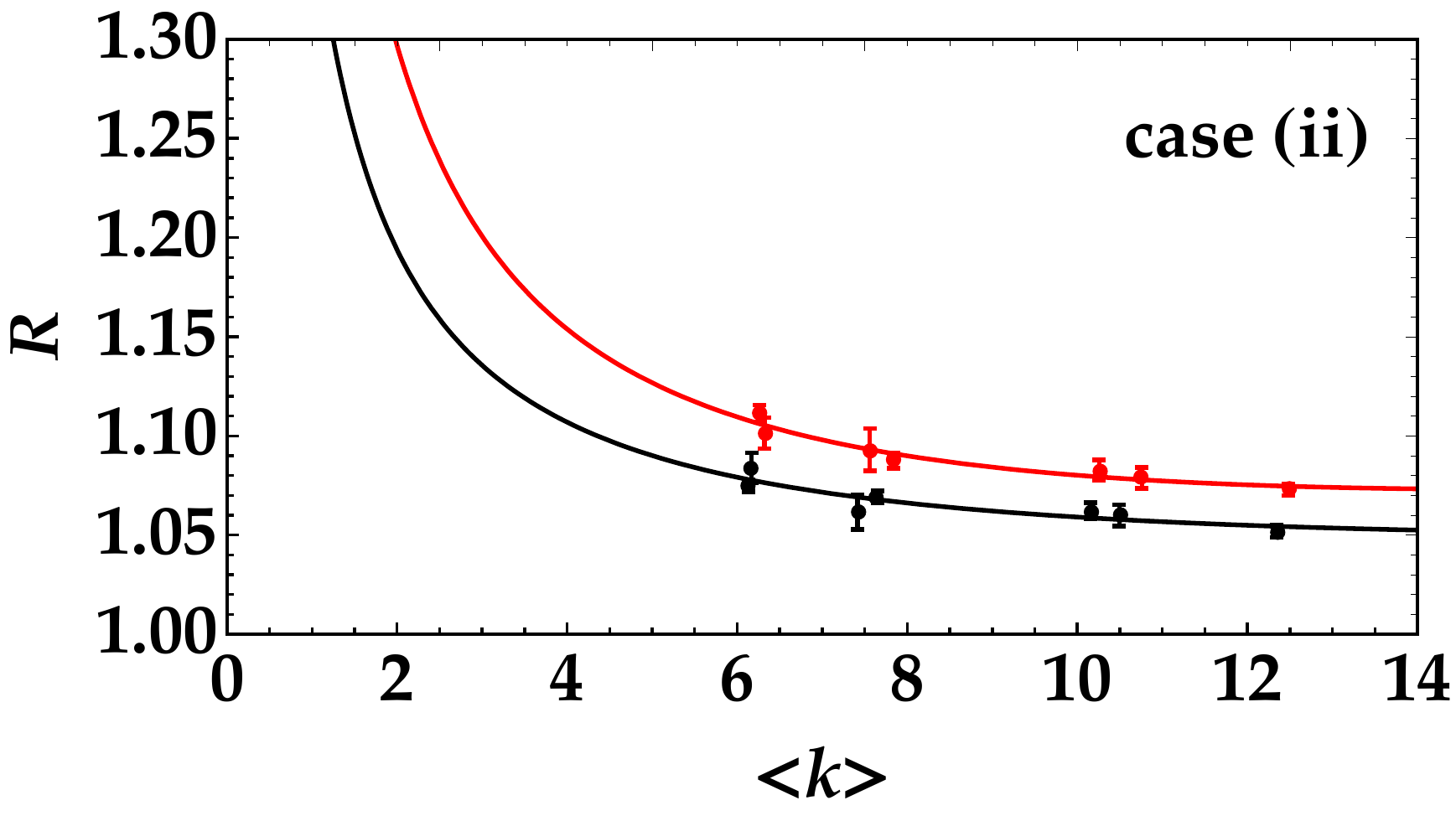}}
\caption{Noise reduction factor for the single-mode thermal state as a function of the mean number of photons detected at a BS output. Colored dots and error bars: experimental data obtained with the system PSAU ($g = 10$~dB) and DRS4 digitizer integrating the traces over $\tau=70$~ns (black dots) and $\tau =110$~ns (red dots); colored lines: theoretical expectations according to Eq.~(\ref{eq:RtotTWBt}). The fitting parameters for $\tau=70$~ns are: $\epsilon = 0.026$, $\langle m_{\rm 1dc}\rangle = 1.406$, $\langle m_{\rm 2dc}\rangle = 0.594$, $t=0.958$ and $\chi^{(2)}=1.5$, while those for $\tau = 110$~ns are: $\epsilon = 0.038$, $\langle m_{\rm 1dc}\rangle= 1.515$, $\langle m_{\rm 2dc}\rangle = 0.485$, $t=0.930$ and $\chi^{(2)}=1.2$.}
\label{R_therm7feb}
\end{figure}

\begin{figure}
\resizebox{0.45\textwidth}{!}{\includegraphics{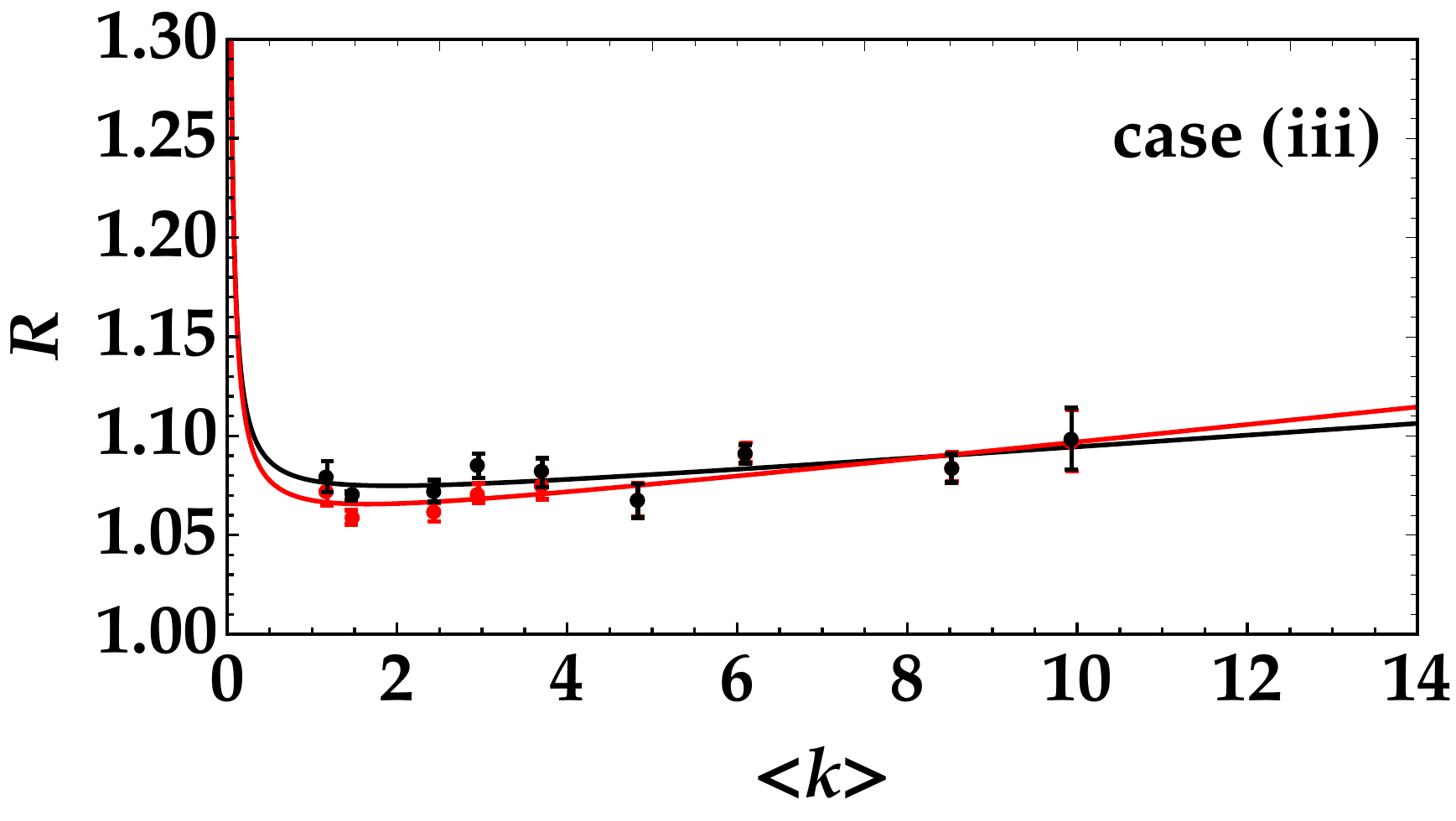}}
\caption{Noise reduction factor for the single-mode thermal state as a function of the mean number of photons detected at a BS output. Colored dots and error bars: experimental data obtained with the slow amplifier and DRS4 digitizer by selecting the peak values (black dots) or integrating over $\tau = 2$~ns before the peak (red dots); colored lines: theoretical expectations according to Eq.~(\ref{eq:RtotTWBt}). The fitting parameters for the peak are: $\epsilon = 0.026$, $\langle m_{\rm 1dc}\rangle = 0.278$, $\langle m_{\rm 2dc}\rangle= 0.422$, $t=0.924$ and $\chi^{(2)}=3.0$, while those for $\tau = 2$~ns are: $\epsilon = 0.018$, $\langle m_{\rm 1dc}\rangle = 0.275$, $\langle m_{\rm 2dc}\rangle = 0.425$, $t=0.907$ and $\chi^{(2)}=2.6$.}
\label{R_therm14feb}
\end{figure}

\noindent
To derive an analytical expression for $R$ that takes all the real effects into account, we exploit the calculation of the correlation function \cite{jointdiff} and implement the detector model introduced in Ref.~\cite{JOSABramilli}. The resulting general expression is \cite{OL19}
\footnote{Note that Eq.~(2) in Ref.~\cite{OL19} contains a misprint: the factor $\langle k_1 \rangle + \langle k_2 \rangle$ should be added as the denominator of the last term.}
\begin{eqnarray}\label{eq:RtotTWB}
R &=& 1 + \frac{1}{\mu}\frac{(\langle k_1 \rangle - \langle k_2 \rangle)^2}{\langle k_1 \rangle + \langle k_2 \rangle} \nonumber\\
&+&\frac{2 \epsilon_1}{1+ \epsilon_1}  \frac{\langle k_1 \rangle}{\langle k_1 \rangle + \langle k_2 \rangle} + \frac{2 \epsilon_2}{1+ \epsilon_2} \frac{\langle k_2 \rangle}{\langle k_1 \rangle + \langle k_2 \rangle}\nonumber \\
&-& \frac{2}{\mu} \left[ (1+ \epsilon_1)\langle m_{\rm 1dc} \rangle - (1+ \epsilon_2)\langle m_{\rm 2dc} \rangle \right]\frac{\langle k_1 \rangle - \langle k_2\rangle}{\langle k_1 \rangle + \langle k_2 \rangle} \nonumber\\
&+&\frac{1}{\mu} \frac{\left[ (1+ \epsilon_1) \langle m_{\rm 1dc} \rangle - (1+ \epsilon_2) \langle m_{\rm 2dc} \rangle \right]^2}{\langle k_1 \rangle + \langle k_2 \rangle}\nonumber\\
&-& 2\sqrt{\eta_1 \eta_2 } \sqrt{\frac{(1+\epsilon_1)\left[\langle k_1 \rangle- (1+\epsilon_1) \langle m_{\rm 1dc}\rangle\right]}{\langle k_1 \rangle + \langle k_2 \rangle}}\nonumber\\
&\ & \ \ \ \ \ \ \times \ \sqrt{\frac{(1+\epsilon_2) \left[\langle k_2 \rangle- (1+\epsilon_2) \langle m_{\rm 2dc}\rangle\right]}{\langle k_1 \rangle + \langle k_2 \rangle}},
\end{eqnarray}
where $\mu$ is the number of multi-thermal modes, $\eta_j$ is the detection efficiency of the detection chains in the two arms, $\langle k_j \rangle = (\eta_j\langle n _j\rangle + \langle m_{\rm jdc}\rangle)(1+\epsilon_j)$ is the mean value of the detector output, $\langle m_{\rm jdc}\rangle$ the mean value of dark counts and $\epsilon_j$ the cross-talk probability. According to the model in Ref.~\cite{JOSABramilli}, the dark-count probability distribution is assumed to be Poissonian, so that Eq.~(\ref{eq:RtotTWB}) also describes the case in which some stray light is detected simultaneously with the signal. Note that the last term vanishes for classically correlated light \cite{highorder}.\\
Since, in spite of all the preliminary alignment procedure, it is not possible to exclude the presence of an imbalance between the detected photons in the two arms of the bipartite state, we introduce the imbalance coefficient $t \in [0,1]$ defined so that $\langle k_1 \rangle \equiv \langle k \rangle$ and $\langle k_2 \rangle = t \langle k \rangle$ \cite{JOSABramilli}. Under this assumption, Eq.~(\ref{eq:RtotTWB}) simplifies to:
\begin{eqnarray}\label{eq:RtotTWBt}
R &=& 1 + \frac{1}{\mu}\frac{(1-t)^2}{1+t}\langle k \rangle +\frac{2}{1+t} \left(\frac{\epsilon_1}{1+ \epsilon_1}+\frac{\epsilon_2}{1+ \epsilon_2}t\right)\nonumber \\
&-& \frac{2}{\mu} \frac{1-t}{1+t}\left[ (1+ \epsilon_1)\langle m_{\rm 1dc} \rangle - (1+ \epsilon_2)\langle m_{\rm 2dc} \rangle \right] \nonumber\\
&+&\frac{1}{\mu} \frac{1}{(1+t)\langle k\rangle}\left[ (1+ \epsilon_1) \langle m_{\rm 1dc} \rangle - (1+ \epsilon_2) \langle m_{\rm 2dc} \rangle \right]^2\nonumber\\
&-& 2\frac{\sqrt{\eta_1 \eta_2 }}{1+t} \sqrt{(1+\epsilon_1)\left[1- (1+\epsilon_1) \frac{\langle m_{\rm 1dc}\rangle}{\langle k \rangle}\right]}\nonumber\\
&\ & \ \ \ \ \ \ \times \ \ \sqrt{(1+\epsilon_2)\left[t- (1+\epsilon_2) \frac{\langle m_{\rm 2dc}\rangle}{\langle k \rangle}\right]}.
\end{eqnarray}
To test the model we start considering classically-correlated light states, namely a single-mode pseudo-thermal state divided at a beam splitter, and evaluate the noise reduction factor from data acquired with three different detection chains.
In more detail, we consider SiPMs followed by the three different amplification and acquisition chains introduced above: (i) PSAU+DT5720 ($g = 12$~dB), (ii) PSAU+DRS4 ($g = 10$~dB) and (iii) slow amplifier+DRS4. For each case, we also investigate the role played by the integration gate width by comparing the results achieved with two different values of $\tau$.
The experimental data are shown in Fig.s~\ref{R_therm10jan}, \ref{R_therm7feb} and \ref{R_therm14feb}, respectively. In each figure, the results corresponding to the shortest $\tau$ are shown as black dots + error bars, while those corresponding to the longest one are shown as red dots + error bars. In the same figures, the theoretical expectations are presented as colored curves with the same color choice.\\
We notice that the experimental values of $R$ are always larger than 1 and that the largest values are obtained in Fig.~\ref{R_therm10jan}, that is the case of acquisition with the Caen system PSAU+DT5720. In this case, the two datasets refer to $\tau=48$~ns (black dots) and $\tau = 80$~ns (red dots). The theoretical expectations were obtained according to Eq.~(\ref{eq:RtotTWBt}), in which we set the number of modes $\mu = 1$, and left the cross-talk probability, the dark-counts and the imbalance as free parameters. In particular, we assumed $\epsilon_1 = \epsilon_2$, and $\langle m_{\rm 1dc} \rangle \neq \langle m_{\rm 2dc} \rangle$.
By looking at the fitting parameters, we can notice that the value of the cross-talk probability is compatible with that indicated in the datasheets and that, as expected, the fitting value slightly increases at increasing the gate width. Moreover, in both cases the two BS outputs exhibit a quite different dark-count contribution, whereas the imbalance coefficient is very close to 1.\\
A similar behavior is achieved for the case (ii) with the second acquisition chain, that is based on the amplifiers of Caen Unit ($g = 10$~dB) and the DRS4 digitizer. The experimental values of $R$ are shown in Fig.~\ref{R_therm7feb} as colored dots + error bars, whereas the theoretical models according to Eq.~(\ref{eq:RtotTWBt}) are plotted as colored curves with the same color choice. In particular, the two datasets refer to $\tau = 70$~ns (black dots) and $\tau = 110$~ns (red dots). The fitting procedure yields values of cross-talk comparable with the data in Fig.~\ref{R_therm10jan}. As to the dark-count contribution, we obtain different values in the two arms, as expected from the decreasing trend of the data. We notice that, according to the values of the reduced $\chi^{(2)}$, the best fit is obtained for $\tau=110$~ns, that is by integrating the entire signal.
%
\noindent
For the case (iii), based on the shaping amplifiers and the DRS4 digitizer, the situation is definitely different. The data are shown as colored dots and error bars in Fig.~\ref{R_therm14feb} together with the theoretical expectations according to Eq.~(\ref{eq:RtotTWBt}) with the same color choice. The black dots are obtained by selecting the peak values, whereas the red dots correspond to the integral of the signal over $\tau=2$~ns before the peak value.
%
In this case, the behavior of $R$ as a function of the mean values exhibits a weak increasing trend. For what concerns the cross-talk probability, we notice that the smallest value is obtained for the integration over $\tau =2$~ns. This is not surprising since integrating the signal amplified by the slow amplifier over 2~ns (10 points) is equivalent to a smoothing operation that reduces possible irregularities given by the simple peak selection. The obtained cross-talk probability is in this case very small, thus proving that the acquisition of the smoothed peak is the best solution among those considered so far.\\
%
%
\noindent
As a general comment on the estimated values of dark counts, we can safely assess that they are larger than the expected values from the sensor datasheet. In fact, at room temperature, the maximum dark-count contribution can be evaluated as $\langle m_{\rm dc}\rangle = 270$~kHz$\times 110$~ns~$=0.029$. To account for the experimental results we can assume the presence of some residual Poissonian infrared light, that is the fundamental of the laser, we could not eliminate completely.

\noindent
To further check the validity of solution (iii), we consider the more interesting case of quantum-correlated optical states, namely the multi-mode twin-beam states.

\subsection{Quantum correlations}\label{sec:TWB}
\begin{figure}
\resizebox{0.45\textwidth}{!}{\includegraphics{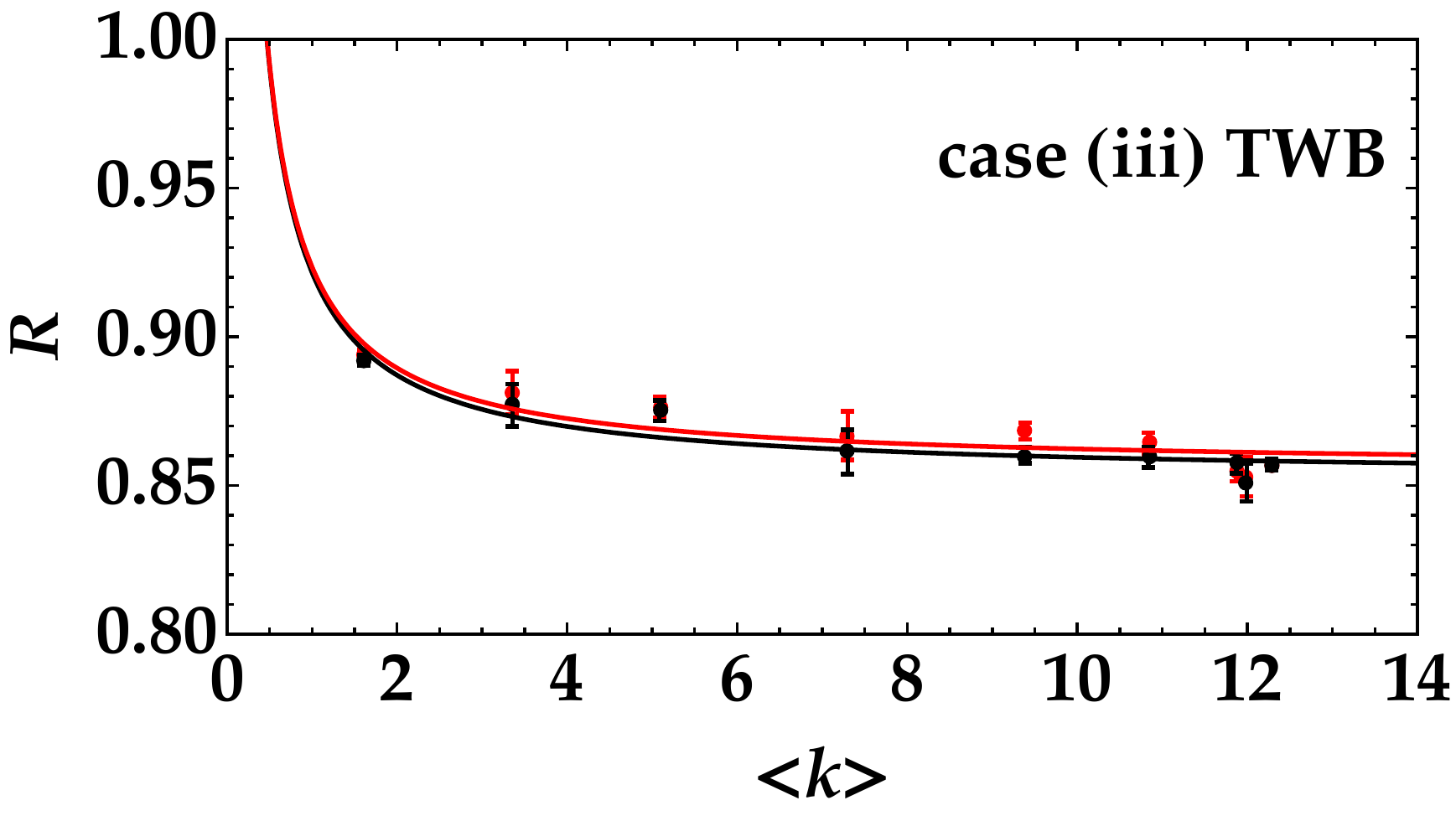}}
\caption{Noise reduction factor for a multi-mode twin-beam state as a function of the mean number of photons on one of the arms. Colored dots and error bars: experimental data obtained with the slow amplifier and DRS4 digitizer by selecting the peak values (black dots) or integrating over $\tau = 2$~ns before the peak (red dots); colored lines: theoretical expectations according to Eq.~(\ref{eq:RtotTWBt}). The fitting parameters for the peak are: $\eta = 0.182$, $\mu = 9256$, $\epsilon = 0.019$, $\langle m_{\rm dc}\rangle=0.349$, $t=0.913$ and $\chi^{(2)}=3.2$, while those for $\tau = 2$~ns are: $\eta = 0.181$, $\mu = 9990$, $\epsilon = 0.020$, $\langle m_{\rm dc}\rangle=0.348$, $t=0.924$ and $\chi^{(2)}=6.2$.} \label{fig:R_twb}
\end{figure}

As anticipated in Section~\ref{sec:lightsources}, we generated multi-mode twin-beam states by sending the fourth harmonic of a Nd:YLF laser to BBO2 of Fig.~\ref{fig:setup}(b) to produce parametric downconversion. \\
For a fair comparison with the classical case of light discussed in the previous Section, we consider the noise reduction factor in Eq.~(\ref{eq:RtotTWBt}). We expect $R<1$ for nonclassical light.
The experimental values of $R$, calculated according to Eq.~(\ref{eq:Rexp}), are shown as colored dots and error bars as a function of the mean number of photons detected in the two arms. The black dots are obtained by selecting the peak values, whereas the red dots correspond to the integral of the signal over $\tau=2$~ns before the peak value.
The fitting procedure yields a low value of cross-talk probability and a non-negligible dark-count contribution that is the same in the two arms. The estimated values of cross-talk probability are very close to those for classical light case (iii) 2~ns gate (see captions of Fig.s~\ref{R_therm14feb} and \ref{fig:R_twb}). Note that, at variance with the case of single-mode pseudo-thermal light, for the multi-mode twin-beam states the difference between selecting the peak and smoothing over 2~ns is less evident, probably due to reduced fluctuations of the multi-mode twin-beam statistics. As in the case of classical light, the estimated dark count values are larger than expected due to the presence of spurious infrared light.

%

%
Finally, we emphasize that the data presented in Fig.~\ref{fig:R_twb} are of better quality than those already reported in Ref.~\cite{OL19}: comparing the same kind of acquisition chains, we see that the mean number of photons is definitely larger (up to 11 instead of 3.5) and the absolute values of $R$ are smaller (0.86 instead of 0.9).
\section{Discussion and conclusions} \label{sec:discussion}

Let us summarize the results presented in the previous Sections and draw some conclusions.

With the aim of detecting and properly characterizing mesoscopic optical states, namely light states endowed with sizeable numbers of photons, by means of SiPMs, we explored and compared different devices, obtained by combining two different kinds of amplifiers and two different digitizers.
In order to choose the best detection chain, we proceeded in two steps. First of all, we analyzed the quality of the pulse-height spectra, introducing the visibility $v$ as a figure of merit for the quality of the spectrum. The analysis of the spectra taken for different integration gate widths indicates that, in the case of well-populated light states, to extract the proper amount of information from the measurements requires the integration of the entire output trace. However, this choice does not exclude all the drawbacks affecting SiPMs since also dark counts, cross talk and afterpulses are acquired together with the light signal.\\
To reduce the incidence of drawbacks, a different strategy can be implemented, consisting in acquiring only a portion of the rising edge of the signal output. Moreover, we have demonstrated that integrating only a small portion of the area under the rising edge or selecting the peak height are equivalent, even if the latter solution is definitely simpler and allows a complete rejection of spurious effects. This is why we devised a proper detection chain in order to correctly catch the peak. The solution addressed in this paper is based on a  shaping amplifier directly connected to the SiPM output and a fast digitizer.\\
In the second part of the paper, we analyzed the problem of light acquisition from the different perspective of getting the correct number of photons shot-by-shot in order to calculate the noise reduction factor $R$. We measured pseudo-thermal light using the three different acquisition chains: the best results are given by the acquisition of the peak values, for which the values of $R$ are closer to unity.
The same acquisition strategy results optimal also in the case of the twin-beam states, which are quantum correlated.\\
In conclusion, the performed investigation leads us to conclude that, provided a proper shaping of the amplified signal and its fast enough digitalization, the strategy of selecting the peak value is the most reliable and even the simplest. On the one side, it avoids the effect of drawbacks because only spurious effects simultaneous to the light signal are acquired, on the other side it holds for any mean photon number unless saturation effects of the detection chain occur. Increasing the dynamic range of the detectors is important to avoid saturation. \\
Finally, we emphasize that the chosen chain is rather compact and can be made portable for possible applications in open air, such as for Quantum Communication.

\section{Acknowledgements}
We acknowledge Giovanni Chesi for useful discussions.

\section{Funding}
No funding. 

\section{Abbreviations}
SiPM = Silicon photomultiplier, PSAU = Power Supply and Amplification Unit, BS = beam splitter, BBO = $\beta$-barium-borate crystal.

\section{Availability of data and materials}
The datasets used and analysed during the current study are available from the corresponding author on reasonable request.

\section{Competing interests}
The authors declare that they have no competing interests.

\section{Authors contributions}
SC, AA and MB conceptualized the work, VM, MP and EV designed the detection chain to properly detect the peak, SC and AA performed the measurements, SC and VM wrote a new code to prepare data for analysis, AA and MB analysed and interpreted the data, AA and MB drafted the work, SC, VM, MP and EV substantively revised it.
All the authors have read and approved the final manuscript.
%
%

\end{document}